# A pyridyl-benzimidazole based ruthenium(II) complex as optical sensor: Targeted cyanide detection and live cell imaging applications

Sudhanshu Naithani,[a] Franck Thetiot,[b] Vikas Yadav,[c] Saakshi Saini,[d] Partha Roy,[d] Samar Layek,*[e] Tapas Goswami*[a], Sushil Kumar*[a]

[a]*Department of Chemistry, Applied Science Cluster, University of Petroleum & Energy Studies (UPES), Dehradun-248007, Uttarakhand, India. (Email: sushil.k@ddn.upes.ac.in; tgoswami@ddn.upes.ac.in).*

[b]*CEMCA, CNRS, UMR 6521, Université de Bretagne Occidentale, Brest 29238, France.*

[c]*Nanoscopic Imaging and Sensing Lab, Indian Institute of Technology Delhi, Hauz Khas, New Delhi 110016, India.*

[d]*Department of Biosciences and Bioengineering, Indian Institute of Technology Roorkee 247667, Uttarakhand, India.*

[e]*Department of Physics, School of Engineering (Applied Science Cluster), University of Petroleum & Energy Studies (UPES), Dehradun-248007, Uttarakhand, India.*

**Abstract**

An extreme toxicity of cyanide ($CN^-$) ion in diverse environmental media has encouraged significant attention for scheming well-organised molecular probes for its selective and sensitive detection. Keeping in mind, we present here a monometallic Ru(II) complex (**Ru-1**) based on 2-(pyridin-2-yl)-1*H*-benzo[*d*]imidazole moiety acting as a highly selective luminescent probe for $CN^-$ recognition in pure water. Besides, **Ru-1** also acted as an efficient sensor for $F^-$, $AcO^-$ and $H_2PO_4^-$ ions along with $CN^-$ when acetonitrile was chosen as solvent system. The binding constant ($K_b$) and detection limit (LoD) for $CN^-$ have been depicted as $3.05 \times 10^6$ $M^{-1}$ and 12.8 nM, respectively, in water. The close proximity of N-H site with Ru(II) centre along with its remarkable acidity were identified as mainly responsible for the high selectivity of **Ru-1** toward $CN^-$ in water. Job's plots and DFT analyses were carried out to support the anion binding mechanism. Furthermore, the time-resolved fluorescence (TRF) spectroscopy was performed to assess the cyanide-induced emission lifetime change of **Ru-1** in aqueous medium. In order to investigate applied potential, the probe **Ru-1** was notably developed into paper-based strips that could readily detect $CN^-$ ion in mM range *via* naked eye under 365 nm light illumination, and also adequately employed to detect $CN^-$ in human breast cancer MCF-7 cell lines and natural food sources (such as apple seeds and sprouting potatoes).

## Introduction

The development of compounds that can selectively detect anions is on high demand as they play a significant role in numerous biological, environmental, industrial, and chemical processes [1–4]. However, this is quite challenging due to the varied shapes, sizes, and high solvation features of anions [5,6]. Among the various anions, cyanide ($CN^-$) is of singular relevance as it is an extremely toxic inorganic anion that results into detrimental effects on aquatic ecosystems and human health. Indeed, it is fatal in small concentrations, and can be easily absorbed by cutaneous, respiratory as well as digestive pathways [7–9]. For instance, cyanide ion interrupts aerobic respiration by inhibiting the cytochrome *c* oxidase; it hinders the production of ATP (i.e., adenosine triphosphate) from cells with a dysfunction of the electron transport chain [10]. Consequently, the World Health Organisation (WHO) has set the maximum permissible limit for cyanide as 1.9 μM in drinking water. Despite its high toxicity, cyanide-based compounds are widely used in various fields such as electroplating, artificial-plastic and gum industries, X-ray film recovery, gold-silver leaching, as well as in the development of diverse range of organic chemicals, pharmaceuticals, and polymers [11–15]. Cyanide contamination and poisoning may result from many sources including industrial exposures, structural fires, medical exposures (e.g., sodium nitroprusside), and some foods (e.g., apple seeds, sprouting potatoes, lima beans and almonds) [16]. Accidental discharge and the subsequent leakage of $CN^-$ from industrially produced wastewater into the domestic water is quite inevitable. *In fine*, water arose as the key medium of biological and environmental systems accountable for major trafficking of the cyanide contamination. Accordingly, it becomes critical to specifically design sensors that can be directly used for $CN^-$ detection in pure water instead of other organic solvent systems. Among the various conventional detection methods such as chromatography, spectrophotometry, flow injection and electrochemical analysis [17–19], the optical (luminescent and/or chromogenic) sensors are emerging as viable alternatives for anion detection due to their archetypal simplicity, satisfying selectivity, high sensitivity, on-site detection, and instant response time [20–26].

In recent years, several cyanide-responsive optical sensors having different recognition units (e.g., amide, benzyl, urea, thiourea, imidazolium, indolium, borane, boronic acid, oxazine, coumarin and hydrazone derivatives) have been reported [27–32]. So far, two major approaches have been typically employed to design a $CN^-$ responsive optical probe. The most common is the chemodosimeter (or reaction-based) sensing approach in which $CN^-$ attacks an electrophilic >C=O (aldehyde or ketone) group or binds with a metal centre to elicit an obvious change in absorption/emission spectra of the sensing probe [33–35]. However, such sensors commonly function in either organic or organic-aqueous solvent systems; *a contrario*, detection of this non-innocent ion in 100% aqueous medium is rather challenging and less explored. The reason lies notably behind the high solvation of $CN^-$ in water depressing the extent of carbonyl-$CN^-$ interaction. Such a drawback limits the applications of these cyanide-responsive chemodosimeters in real-life samples [36].

The second approach combines an anion binding site with luminophores (or signalling unit) through a $\pi$-conjugated spacer. Conventionally, a $CN^-$ ion interacts with binding site, often *via* H-bonding (hydrogen-bonding), and triggers the detectable optical spectral changes in the probe [15,37–39]. The latter has been essentially used to recognize acetate ($OAc^-$) and halide ions; however, the H-bonded cyanide-responsive probes, particularly functioning under physiological conditions, are still less explored [40,41]. This is due to the fact that cyanide ion often loses its charge at physiological pH through involvement in protonation equilibria; and thus, it exhibits less affinity toward weak H-bonding donors [42]. Nonetheless, as compared to $F^-$ and $OAc^-$ ions, $CN^-$ acts as a stronger base in water. These observations cohesively hint that the acidic character of H-bonding donors plays an important role in selective detection of $CN^-$ among other competing anions (e.g., $F^-$ and $OAc^-$) in water [43]. Besides, our group highlighted in a recent study the significance of imidazo-phenanthroline based luminescent probes having the N-H functionality as the key site for anion binding and recognition [44]; the anion binding affinity of such probes could easily be tuned by adjusting the acidic character of N-H site. The stronger acidity of the latter results into stronger binding of anions. Indeed, recent development in this field has revealed that the probes having multipoint H-bonding may lead to high anion affinity, ultimately enabling the $CN^-$ responsive probes to tolerate an extensive amount of water from the solvent [43].

In this context, ruthenium(II)-polypyridyl based luminescent probes are receiving wide attention due notably to their excellent photophysical properties, large stokes shift, appreciable water solubility, long excited-state lifetime and visible light emission/excitation [45–49]. Resultantly, several mono- and bi-metallic Ru(II) probes have been reported for the detection of cyanide *via* both binding- and reaction-based approaches [6,43,50–57]. Though significant achievements have been obtained for optical cyanide sensing, some disadvantages including sensor's reversibility and lack of efficiency with higher LoDs (limit of detection) in protic media cannot be ignored. In addition, many of such sensors suffer from the complicated multi-step synthetic procedures, poor solubility and low analyte selectivity in water. Therefore, the development of simple architectures, cost-effective, highly selective/sensitive cyanide-responsive probes is essential and remains a challenge.

In this endeavour, we demonstrate an easily synthesizable Ru(II)-based optical probe (**Ru-1**) derived from 2-(pyridin-2-yl)-1*H*-benzo[*d*]imidazole ligand for the selective detection of $CN^-$ ion particularly in water (Scheme 1). Probe **Ru-1** has been prepared earlier [58,59]; however, to the best of our knowledge, it was never used for anion sensing purpose. Thus, we anticipately and rationally hypothesized that the sensing ability of **Ru-1** would be more pronounced in comparison to the previously reported systems because the N-H group is in relative proximity of the metal centre in **Ru-1**; as a result, the N-H proton of the imidazole moiety becomes substantially acidic due to the strong electron pull of Ru(II) centre. Beyond the selective detection of $CN^-$ in aqueous media, probe **Ru-1** has also been employed in organic as well as in aqueous-organic media to recognise other anions further defining the versatility of the probe; hence, it could detect $F^-$, $AcO^-$, and $CN^-$ ions in acetonitrile ($CH_3CN$) solution *via* H-bonding interaction. The effect of water content on the selectivity of this probe for various anions has also been scrutinized in this report. About the applied potential, the $CN^-$ recognition ability of probe **Ru-1** was purposely tested in water-based real samples such as drinking water, tap water, natural food sources (apple seeds and sprouting potatoes) as well as in MCF-7 live cells. Beyond, this probe proved to be of relevance to detect $F^-$ and $AcO^-$ ions in fluoride-containing toothpaste/mouthwash and commercial vinegar samples, respectively, in semi-aqueous ($CH_3CN$-$H_2O$) solvent system.

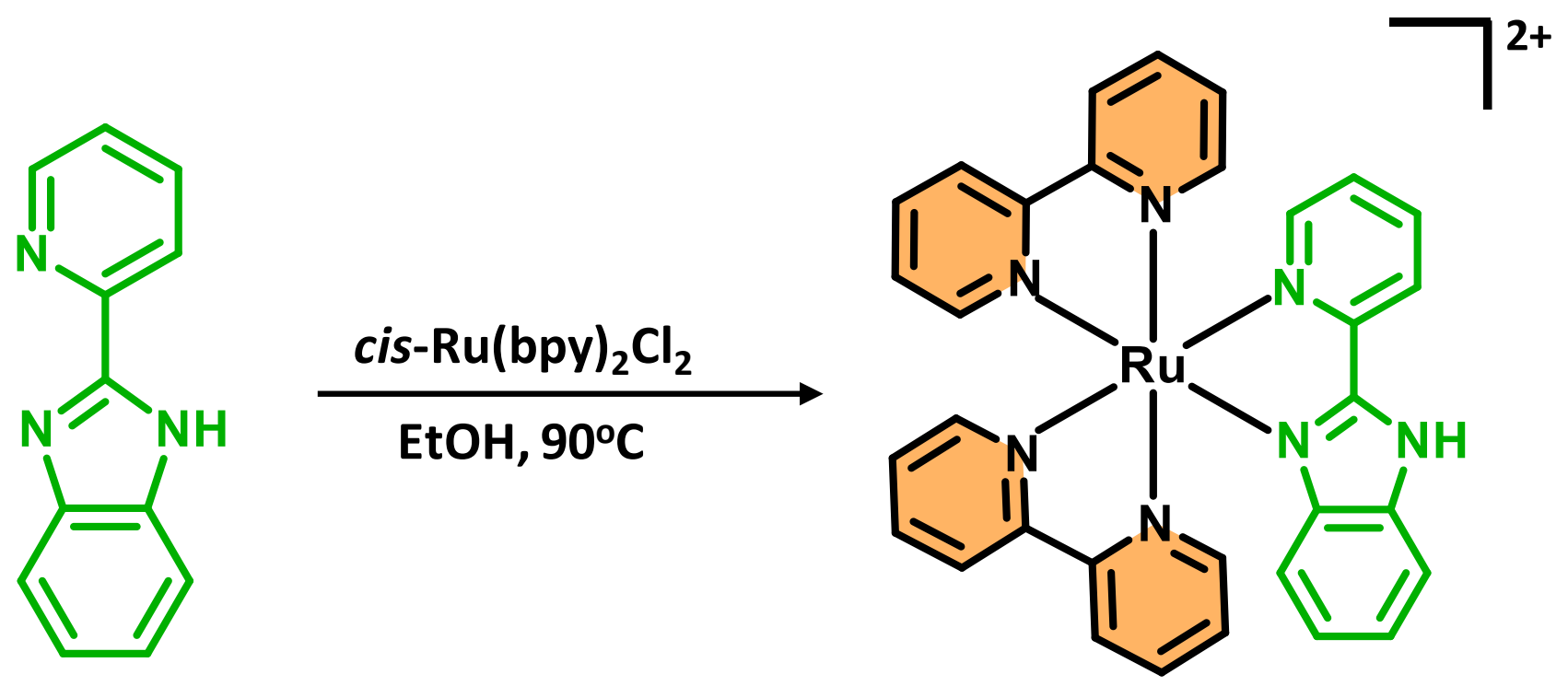

**Scheme 1**. Synthesis of Ru(II)-based luminescent probe **Ru-1**.

## Experimental Section

**Materials and Measurements**. Reagent grade chemicals procured from the commercial sources were utilized as received. Doubly distilled de-ionized water and HPLC grade $CH_3CN$ have been used to perform UV-Vis and luminescence measurements. Elemental (C, H and N) analysis has been measured on Elementar-Vario-EL analyser. ESI-MS analysis for **Ru-1** was performed on Agilent Waters Q-TOF Premier HAB213 mass spectrometer. Electronic absorption spectra have been obtained on Shimadzu UV-240 spectrophotometer, and the emission spectra were recorded on Cary Eclipse Fluorescence spectrophotometer. FT-IR studies were carried out on Shimadzu IR-Spirit using KBr pellets. Emission lifetimes were obtained on FLS920 fluorescence lifetime and steady-state spectrometer using a light source of 500 kHz laser beam at 405 nm.

**Synthesis of Ru-1**. Complex **Ru-1** has been synthesized according to the literature procedure [58,59]. Briefly, a batch of 0.1 mmol *cis*-Ru(bpy)$_2$Cl$_2$, 0.1 mmol 2-(pyridin-2-yl)-1*H*-benzo[*d*]imidazole, and 20 mL ethanol were added to a 50 mL three-neck round bottom flask. The reaction mixture was refluxed at 90°C for 10-12 h under inert atmosphere. The color of the solution turned red orange. The complex was precipitated out by addition of a saturated aqueous solution of sodium hexafluorophosphate ($NaPF_6$) and then collected using vacuum filtration. Ultimately, complex **Ru-1** has been purified using silica-gel column chromatography with $CH_3CN$ as eluent followed by recrystallization from $CH_3CN$-MeOH (1:1, *v/v*) mixture. (Yield: 88 %). Anal. Calcd. for $C_{32}H_{25}F_{12}N_7P_2$: C, 42.77; H, 2.78 and N, 10.91; Found: C, 42.72; H, 2.85

and N, 10.76. FT-IR (KBr disk): 3360, 3064, 1615, 1462, 1240, 962, 838, 760 and 550 $cm^{-1}$. UV-Vis ($H_2O$): $\lambda_{max}$/nm = 285 nm and 460 nm. ESI-MS (positive, $CH_3CN$): *m/z* Calcd. for $[M-H-PF_6]^+$ = 608.11, Found = 608.1110; *m/2z* Calcd. for $[M-2PF_6]^{2+}$ = 304.56, Found = 304.5589.

**Spectroscopic titrations**. The sensing behavior of probe **Ru-1** for different anions was investigated in water and $CH_3CN$ (10.0 μM) solutions. The stock solutions of anions (1.0 mM) were prepared just before performing the sensing experiments using tetrabutylammonium or potassium salts of $F^-$, $Cl^-$, $Br^-$, $H_2PO_4^-$, $CN^-$, $I^-$, $AcO^-$, $NO_3^-$, $SO_3^{2-}$, $S^{2-}$, $SO_4^{2-}$ and $NO_2^-$ ions. Quartz cuvettes of 1.0 cm path length and 3.0 mL volume were used for all the UV-Vis measurements. For a typical UV-Vis titration, 100 μL aliquot of a given anion (1.0 mM) was added to 2900 μL solution of the probe (10.0 μM). For emission titration experiment, 10 μL aliquot of a given anion (1.0 mM) was added to 2990 μL solution of the probe (2.0 μM), and the excitation wavelength ($\lambda_{ex}$) was fixed at 460 nm maintaining the excitation slit width as 5.0 mm.

**Calculation of limit of detection (LoD)**. The detection limits for anions have been determined with the help of emission titration profiles of probe **Ru-1** with various anions (i.e., $F^-$, $AcO^-$, $H_2PO_4^-$ and $CN^-$) using the typical equation LoD = 3σ/S (where σ stands for standard deviation calculated from 10 blank measurements, and S is the slope calculated from the calibration curve) [60].

**Calculation of binding constant ($K_b$)**. The binding constants of **Ru-1** for different anions have been calculated from the emission titration profiles using the modified Benesi–Hildebrand equation as follows:

$$1/(F-F_0) = 1/(F_{max} - F_0) + 1/(F_{max} - F_0)\,(1/K_b)\,(1/[A^-])$$

Where $F_{max}$ is the fluorescence intensity of **Ru-1**; $F_0$ stands for the fluorescence intensity of **Ru-1** with anions at saturation, and F is the fluorescence intensity of **Ru-1** with anion at some intermediate concentration. The $[A^-]$ is the concentration of the anion at a given time. The binding constant is determined from the plot of $1/(F-F_0)$ against $1/[A^-]$ *via* the slope and the intercept of the straight line [61].

**Job's plot analysis**. The absolute concentration of probe **Ru-1** with different anions was kept at 1.5 μM. The mole fraction was then varied from 0 to 1.0 [62]. Finally, a graph was plotted for the emission response of probe **Ru-1** with $[(F-F_0) \times (1-X)]$ and mole fraction (X) of $A^-$ ions.

**Test paper strips experiments**. The low-cost test paper strips experiments were carried out by immersing the finely cut-shaped Whatman filter paper strips into an aqueous solution of **Ru-1** (50 μM). The test strips were then taken out and air-dried. After that, the treated strips were kept into the solutions of various selected anions (*ca.* $10^{-2}$ M) to eventually observe the visual changes under UV as well as visible light illumination.

**Cell culture and imaging.** The human breast cancer cells (MCF-7) were seeded in DMEM-HG (Dulbecco's modified Eagle's medium-High glucose) medium and further supplemented with 10% heat-inactivated fetal bovine serum (FBS). The cells were seeded in a 24-well plate at the density of $0.05 \times 10^6$ cells/well and cultured for 24 h at 37°C with 5% $CO_2$. The MCF-7 cell lines were incubated with 20-100 μM of probe **Ru-1** for 3 h. Thereafter, the cells were washed with PBS to remove excess of **Ru-1**. The cells were then imaged under an inverted fluorescence microscope (EVOS FLOID, ThermoFisher, USA). The fluorescently-labeled cells were then incubated with 2.0-5.0 eq of $CN^-$ ions for 3 h. The cells were again washed with PBS buffer to remove excess cyanide and then imaged under the fluorescent microscope.

**DFT calculations**. The DFT (density functional theory) calculations have been performed for probe **Ru-1** and its deprotonated form (i.e., **Ru-1-dp**) without symmetry constraints by using Gaussian 09W program. The geometries of the complexes have been derived with the help of semi-empirical PM3 level for the molecular frameworks. Moreover, the analytical vibrational frequency was calculated using the long-range corrected (LC) hybrid density functional with empirical distribution correction (GD3B) by calculating ZPE at 273.15 K T and 1.0 atm pressure. The structures were optimized using B3LYP theory with 6-311G++(d,p) basis set for the C, N, and O atom and LANL2DZ basis set for Ru centre with effective core potential. The numerical simulation was performed using a superfine grid.

## Results and Discussion

**Synthesis**. Probe **Ru-1** has been synthesized following the procedure reported earlier [58,59], and notably characterized by UV-Vis, fluorescence, FT-IR, ESI-MS, and elemental analyses.

UV-Vis studies clearly revealed the characteristic Ru(II)-bpy centred bands at 285 nm ($\pi \rightarrow \pi^*$; ligand centred (LC)) and 460 nm ($d\pi_{Ru} \rightarrow \pi^*_{bpy}$; metal-to-ligand-charge-transfer (MLCT)) (Fig. 1(a)) while the emission maximum for **Ru-1** was observed at 625 nm in pure water (Fig. 1(b)). In $CH_3CN$ solution, the absorption and emission maxima appeared at 457 nm and 622 nm, respectively (Fig. S1-S2). The peaks in ESI-MS at 304.5589 and 608.1110 have been attributed to the original (*m/2z*) and deprotonated (*m/z*) molecular ion peaks (Fig. S3).

**Sensing behavior of Ru-1 in aqueous media.** The development of highly water-soluble and luminescent probes remains the need of modern-day-research for their practical utility and successful implementation into real-life samples. Both UV-Vis and emission spectral studies have been employed to assess its sensing ability toward various anions of relevance in 100% aqueous medium. A 10.0 μM solution of **Ru-1** was prepared in deionised water and was tested against various selected anions (such as $F^-$, $Cl^-$, $Br^-$, $H_2PO_4^-$, $I^-$, $CN^-$, $AcO^-$, $NO_3^-$, $SO_3^{2-}$, $S^{2-}$, $SO_4^{2-}$ and $NO_2^-$). As expected, complex **Ru-1** served as a highly selective sensor for $CN^-$ ion in water amongst the pool of tested anions. The progressive addition of $CN^-$ (0-9.0 equiv) resulted in a pronounced red-shift in the MLCT band of **Ru-1** from 460 nm to 487 nm which was accompanied with three isosbestic points at 324 nm, 402 nm, and 477 nm (Fig. 1(c)). The clear isosbestic points suggested the stoichiometric generation of some new species from **Ru-1** upon interaction with $CN^-$ ion. Comparatively, other anions including the competing $F^-$ and $AcO^-$ failed to produce any such changes. These results were further corroborated by emission titration experiments. As shown in Fig. 1(b), the luminescence intensity of **Ru-1** at 625 nm underwent substantial quenching solely in case of $CN^-$ ion (9.0 equiv) over other anions ($\lambda_{ex}$ = 460 nm).

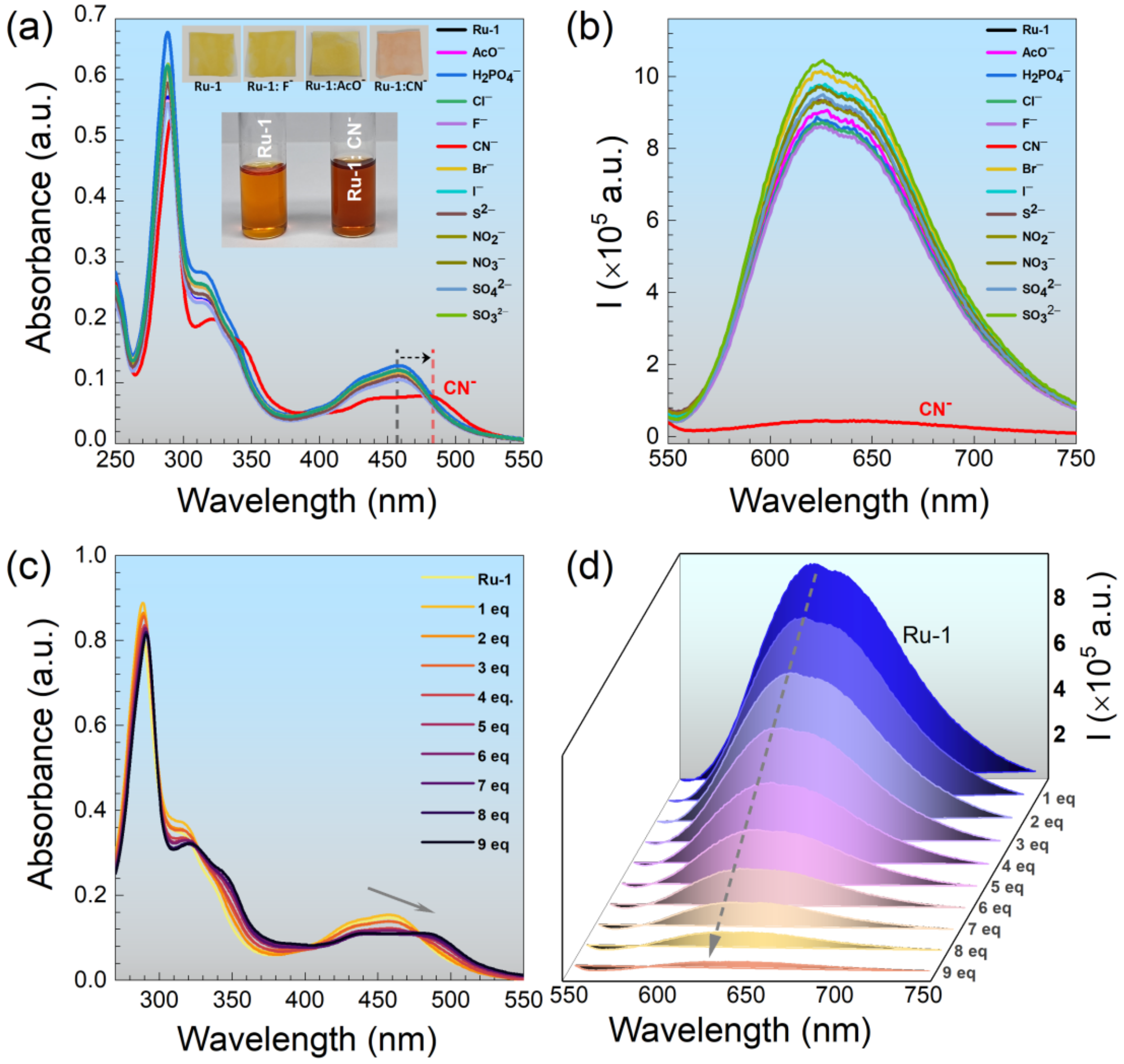

**Fig. 1.** (a-b) UV-Vis and emission spectra of the probe **Ru-1** (*ca.* $10^{-6}$ M) with various anions in water and (c-d) UV-Vis and emission spectral variations in **Ru-1** at different concentrations of cyanide ion (0-9.0 equiv). Inset: Color changes in aqueous solution and on paper strips that occur when **Ru-1** was treated with $CN^-$ ions.

Interestingly, **Ru-1** also acted as a colorimetric probe for $CN^-$ as the original red-orange color of **Ru-1** turned red-brown in presence of $CN^-$ ions (inset of Fig. 1(a)). Job's plot analysis indicated a 1:1 binding stoichiometry for **Ru-1**---$CN^-$ adduct formation (Fig. S4). The limit of detection (LoD) for $CN^-$ has been depicted as 12.8. nM (Fig. 2) which is remarkably lower than the maximum $CN^-$ contaminant level in the drinking water scaled by U.S. EPA (Environmental Protection Agency) [12,41]. To further rationalize the latter, a systematic comparison of the sensing parameters with previously reported $CN^-$-responsive Ru(II)-based luminescent probes has been recapped (see Table 1). The comparison clearly suggests that the relative proximity of imidazole ring with Ru(II) results into an improved LoD value; however, the binding constant was found relatively comparable ranging from $10^5$ to $10^6$ $M^{-1}$ in majority of the cases. Notably, the $CN^-$ addition induced more pronounced MLCT shift in **Ru-1** as compared to the probes integrated with other ligand scaffolds (Table 1).

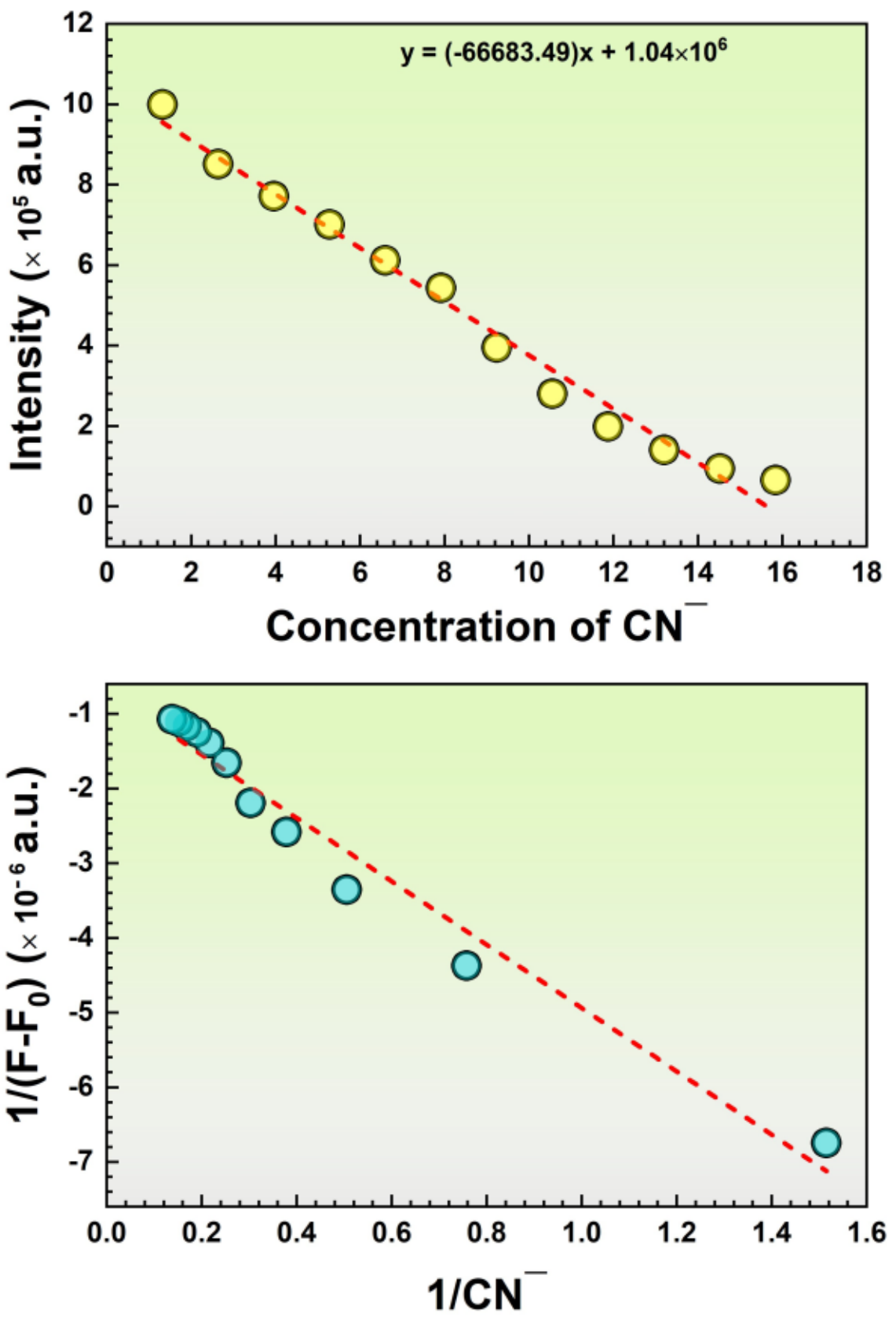

**Fig. 2.** Emission intensity of **Ru-1** at 625 nm *vs*. concentration of $CN^-$ ion, with binding constant calculated from a 1:1 binding model.

**Sensing behavior of Ru-1 in acetonitrile solution**. The anion sensing ability of **Ru-1** was also investigated in $CH_3CN$ solution (Figs. S5-9). A 10.0 µM solution of **Ru-1** was titrated against 6.0 equiv of anionic species including $F^-$, $Cl^-$, $Br^-$, $H_2PO_4^-$, $CN^-$, $I^-$, $AcO^-$, $NO_3^-$, $SO_3^{2-}$, $SO_4^{2-}$ and $NO_2^-$ ions. The MLCT band of **Ru-1** markedly red-shifted (from 457 nm to 497 nm) in the presence of $F^-$, $AcO^-$ and $CN^-$ ions while other anions displayed almost negligible changes (Fig.

S5(a)). The extent of red-shift in acetonitrile (i.e., 40 nm) was more than that observed in water (i.e., 27 nm). Similar to aqueous medium, the absorption changes were associated with three clean isosbestic points at 322 nm, 411 nm and 474 nm. Probe **Ru-1** emitted near 624 nm in $CH_3CN$ when excitation wavelength was 460 nm. The luminescence intensity exhibited a complete quenching upon the addition of $F^-$, $CN^-$, $AcO^-$ and $H_2PO_4^-$ (6.0 equiv) with no remarkable changes in the presence of other investigated anions (Fig. S5(b)). These results clearly revealed that **Ru-1** could be utilized to detect $F^-$, $CN^-$, $AcO^-$ and $H_2PO_4^-$ ions in $CH_3CN$ solution. Interestingly, in $CH_3CN$, the limit of detection LoD was found relatively better (1.30 nM) for $CN^-$ ion when compared with that in aqueous medium (12.8 nM), most likely due to some hydration of cyanide ion in water. Moreover, the LoD values for $F^-$ and $AcO^-$ were depicted as 2.26 nM and 1.90 nM, respectively, in $CH_3CN$ solution.

Distinctly, we also performed the interference experiments in water in the presence of other competing anions. No change in the luminescence of **Ru-1** could be realized when treated with competing anions such as $F^-$ and $AcO^-$ ions even at their high concentrations (upto 10.0 equiv); however, as soon as a small concentration of $CN^-$ was added, the emission band was completely quenched. Hence, the response of **Ru-1** for $CN^-$ was found unaffected by the presence of other assorted anions (Fig. S10), indicating a high selectivity of **Ru-1** for $CN^-$ in aqueous medium. Besides, the probe exhibited reversibility in water in the presence of acetic acid; indeed, the dark-red colored solution of **Ru-1**----$CN^-$ faded, and the original emission/absorption spectra of the probe **Ru-1** restored due to the protonation process induced by acetic acid.

**Effect of water content**. For its practical usage, the changes in UV-Vis spectra of probe **Ru-1** for $F^-$ and $CN^-$ ions were evaluated and monitored in different mixtures of $CH_3CN$-$H_2O$ (80/20, 50/50 and 35/65; v/v), and the results were compared with the data observed in pure $CH_3CN$ solution. In pure $CH_3CN$, nearly 1.2 equiv of either $F^-$ or $CN^-$ completed the binding reaction with a red-shift in the MLCT band from 457 nm to 497 nm. On the other hand, in 20% aqueous solution ($CH_3CN$-$H_2O$, 80/20, v/v), the addition of 3.0 equiv of both the anions was required to exhibit a similar change (Fig. S11(a)). In 50% aqueous solution, *ca.* 20.0 equiv of $F^-$ was required to complete the reaction while the amount used for $CN^-$ was 6.0 equiv (Fig. S11(b)). Further increase into water content (i.e., 65% aqueous solution) resulted into remarkable changes as the reaction was incomplete even after addition of 50 equiv of $F^-$ ions; however, 8.0 equiv $CN^-$ led to the completion of the reaction (Fig. S11(c)). Therefore, the concentration of water is an important parameter for selectivity, and sensitivity of probe **Ru-1**.

Fundamentally, the high selectivity of probe **Ru-1** for $CN^-$ in water may be attributed to a relatively lower hydration value of $CN^-$ ($\Delta G_{cyanide}$ = -295 KJ/mol) as compared to the other anions ($\Delta G_{fluoride}$ = -465 KJ/mol, $\Delta G_{acetate}$ = -365 KJ/mol) [36,43,63]. The other anions basically preferred to be hydrated and did not form any hydrogen bond with the N-H of the imidazole moiety in **Ru-1**. Moreover, $CN^-$ has a larger pKa value (9.0 for HCN) which indicates that it acts as stronger base in water when compared to AcOH (pKa 4.75) and HF (pKa 3.17) that can easily extract a proton. As a result, the high selectivity towards $CN^-$ makes this probe a powerful candidate for the detection of $CN^-$ ion in real aqueous samples.

We have also tested the photosensitivity of probe **Ru-1** with the help of absorption and emission spectral analyses in aqueous medium. Upon irradiation of **Ru-1** with UV and visible light at different time intervals up to 2 hours, the negligible spectral changes could be noticed which clearly indicated the photostable behaviour of our probe (Fig. S12 and S13).

**Time resolved fluorescence (TRF) study**. Luminescence lifetime experiments have also been performed to corroborate the $CN^-$ sensing mechanism. The time resolved emission spectrum of free probe (i.e., **Ru-1**) exhibited a mono-exponential decay, and the emission lifetime was calculated to be 166.9 ns in water (Fig. 3). The gradual addition of $F^-$ and $AcO^-$ displayed no change in its lifetime (Fig. S14); however, the addition of $CN^-$ induced a quenching in the average lifetime from 134 ns to 115 ns (Fig. 3).

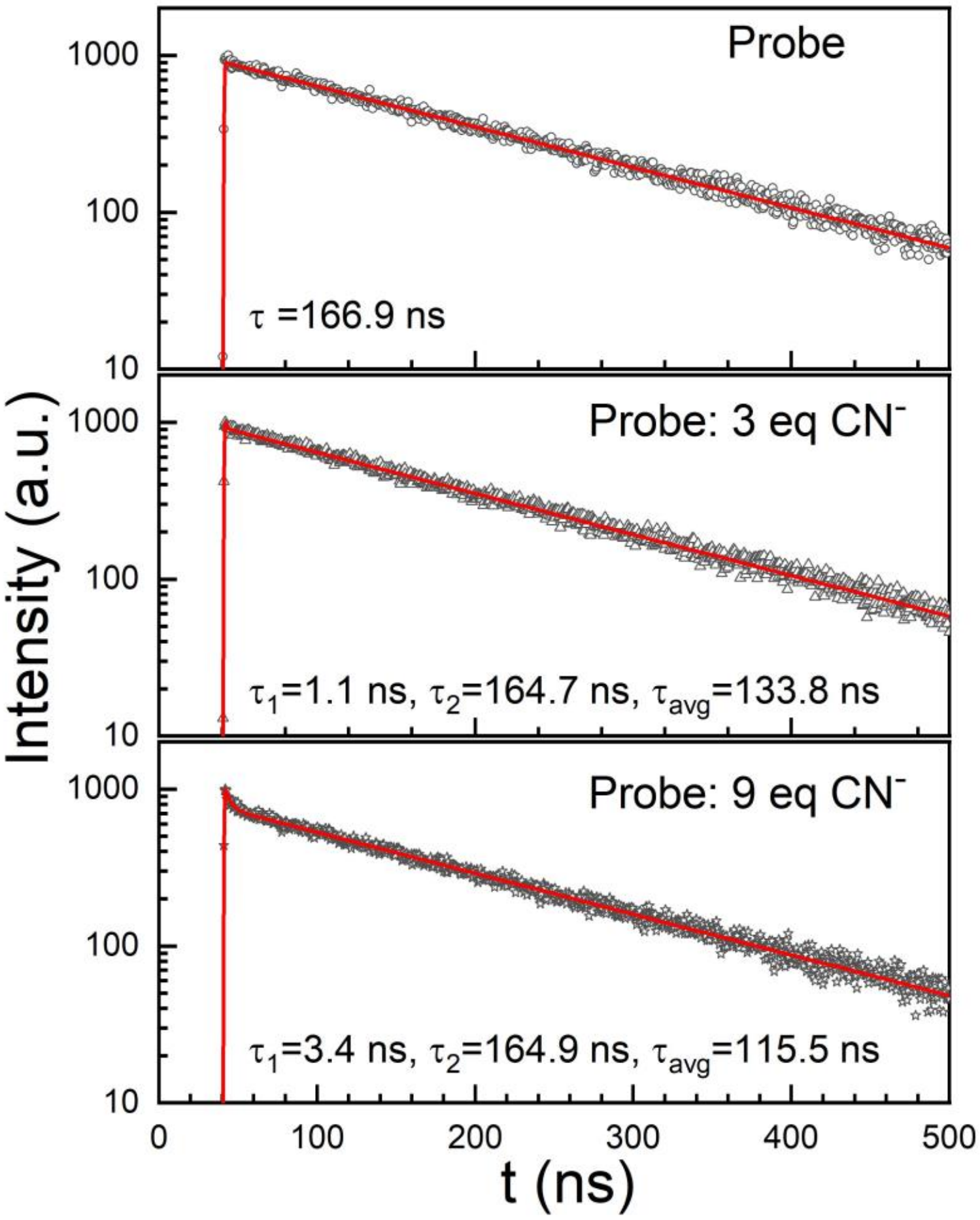

**Fig. 3.** TRF studies of **Ru-1** with different concentrations of $CN^-$ ions.

Notably, unlike the other competing anions, the introduction of $CN^-$ resulted in a bi-exponential decay profile. The concurrent existence of more than one species, each with distinct lifetime, may result in a bi-exponential decay. The rational assumption is that the $CN^-$ ion deprotonates the N-H group making the emission lifetime shorter than the free probe and results in the luminescence quenching. Hence, the observed bi-exponential decay profile might be attributed to the existence of probe **Ru-1** in two states i.e., protonated (i.e., original probe **Ru-1**) and deprotonated (referred as **Ru-1-dp**) having different lifetimes. It is worth noting that the change in lifetime is consistent with the probe's steady-state behavior.

**Detection of anions in real samples**. To further assess its practical relevance and applied potential, the sensing ability of **Ru-1** for $CN^-$ was exploited in real water samples (e.g., drinking water and tap water) and food sources (e.g., apple seeds and sprouting potatoes). The water samples were spiked with varied concentrations of $CN^-$ ions. The detection was monitored by recording the UV-Vis spectral changes. The addition of water sample spiked with $CN^-$ ion to probe's solution induced a red-shift in the MLCT absorption of probe **Ru-1** together with a color change of solution from orange to red-brown (shown in Fig. 4). The cyanide concentration in the spiked water sample could be determined with excellent recovery values measured using UV-Vis and emission spectral analyses (Tables S1 and S2). On the other hand, the luminescence intensity of **Ru-1** declined immediately upon addition of cyanide containing tap water sample to probe **Ru-1** (Fig. S15(a)).

The food samples have been prepared following the procedure reported earlier [64,65]. Apple seeds and sprouting potatoes were crushed and soaked in 50 mL water for a day. The samples were then filtered, washed with NaOH solution and double distilled to adjust a pH of 7.0. As shown in the Fig. S15(b), a significant luminescence quenching of **Ru-1** could be noticed upon the addition of the cyanide containing food samples to the aqueous solution of the probe **Ru-1**. These results validate that **Ru-1** acts a simple, fast and efficient $CN^-$ sensor that can be utilized to detect cyanide in various real samples.

Distinctly, to investigate the versatility of probe **Ru-1**, we have used commercially available vinegar (for $OAc^-$) and mouthwash/toothpaste (for $F^-$) samples. The addition of one drop of either mouthwash or vinegar sample to **Ru-1** led to an instant color change in probe's solution (visible to the naked eye). The $OAc^-/F^-$ sensing ability of **Ru-1** into commercially available mouthwash and vinegar samples was also supported by slightly red-shifted MLCT band at 460 nm in UV-Vis spectra (Fig. 4(a)). For sensing of fluoride ion, the sample was prepared by dissolving 40 mg of toothpaste/mouthwash in 5.0 mL $CH_3CN$ with the help of sonication followed by the centrifugation. The sample solution (30 μL) was then added to the aqueous solution of probe **Ru-1** (10 μL) and the changes were monitored using UV-Vis analysis. A remarkable luminescence quenching in probe **Ru-1** could also be observed upon the addition of the mouthwash and vinegar samples to the acetonitrile solution of probe **Ru-1** (Fig. S15).

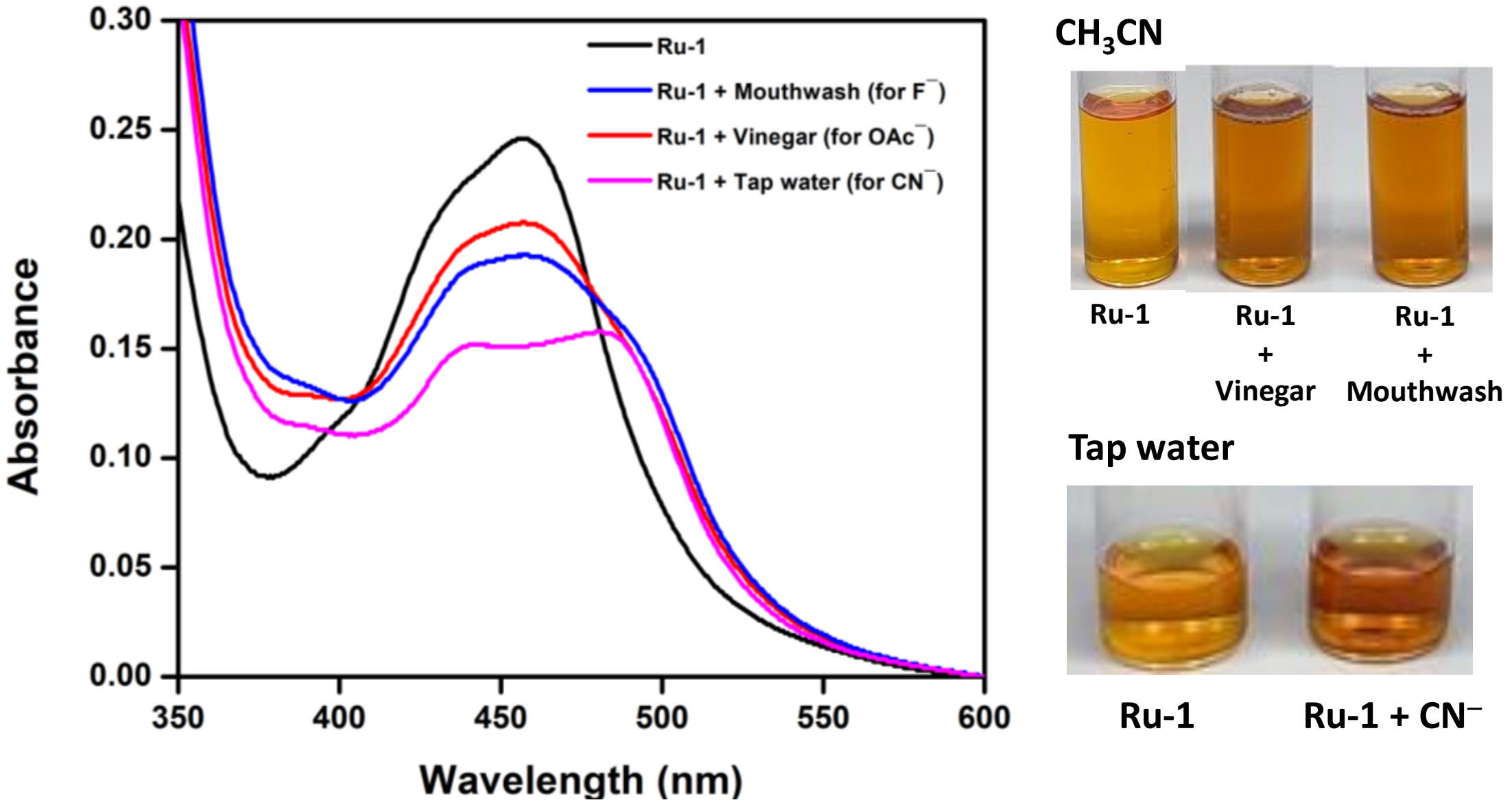

**Fig. 4.** Response of probe **Ru-1** for different real samples. Detection of OAc⁻ in commercially available vinegar (red-line); F⁻ in mouthwash (blue-line) and CN⁻ in cyanide-spiked tap water (pink-line) using UV-Vis spectral analyses.

**Low-cost test strips experiments**. In order to further apply our probe in everyday life applications, we have developed the test strips by soaking the finely cut-shaped filter paper (Whatman Grade) strips in aqueous solution of probe **Ru-1** (*ca.* $10^{-3}$ M) followed by the drying in open air. A significant color change from light yellow to red could be observed when the test strips were treated with CN⁻ ions and visualised under visible light illumination. These results clearly suggested that this probe can be applied to recognise cyanide traces in real samples (inset of Fig. 1(a)).

**Cell imaging study**. The probe **Ru-1** was also investigated for biological application such as cell imaging in human breast cancer MCF-7 cell lines. Upon incubation with variable concentration of **Ru-1** (20-60μM), the cell lines displayed bright red fluorescence as seen in Fig. 5(a). The successive addition of cyanide ions induced significant quenching of luminescence when treated with cells containing 40 μM of probe **Ru-1** (Fig. 5(b)). These results were in consistence with the *in vitro* studies and indicated that the probe **Ru-1** can be utilized for imaging of CN⁻ in biological settings.

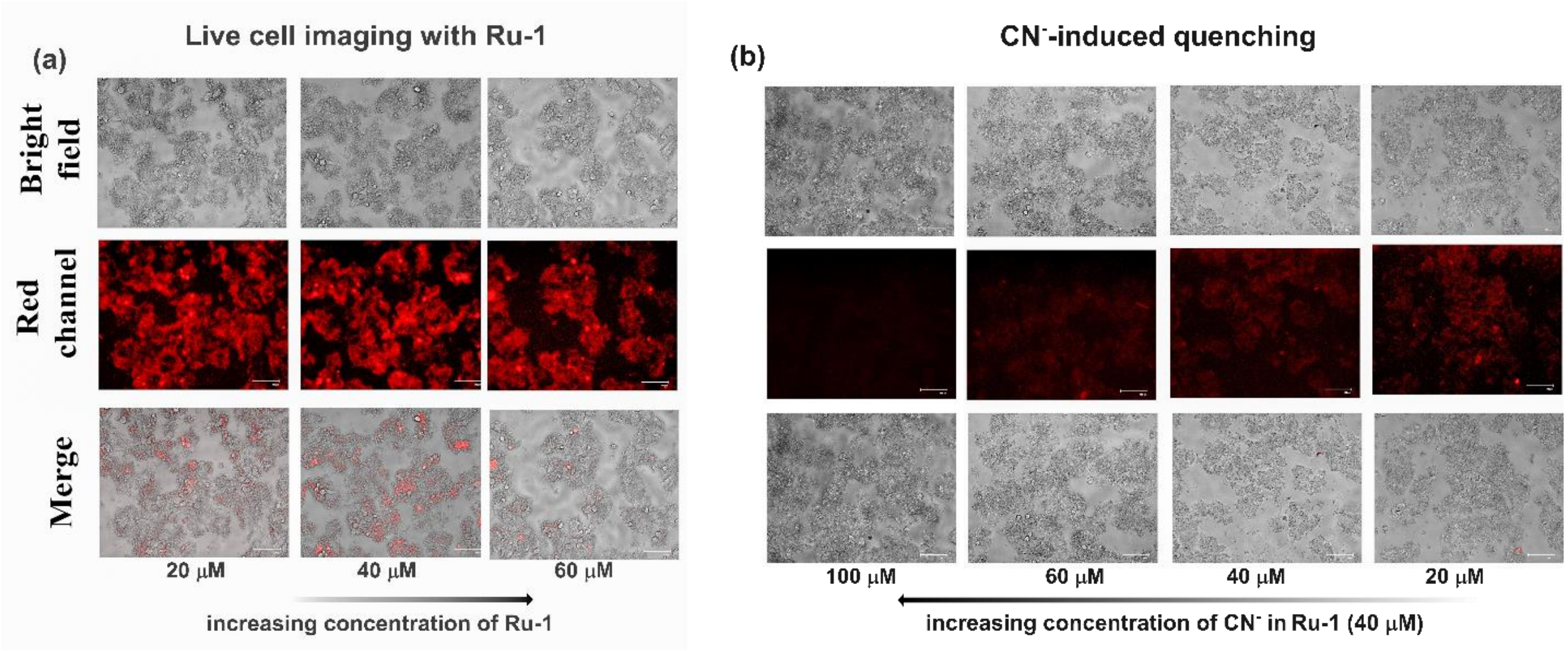

**Fig. 5.** Confocal luminescence imaging of human breast cancer MCF-7 cell lines. (a) Cell lines incubated with variable concentrations (20-60 μM) of only probe **Ru-1** and (b) incubation of cell lines with progressive addition of $CN^-$ ions (20-100 μM) in presence of 40 μM **Ru-1**.

**Mechanism for anion binding**. The outcomes of the experimental findings indicate explicitly that the probe **Ru-1** interacts strongly with $CN^-$ ion in pure aqueous medium. In $CH_3CN$, it acts as a sensor for $CN^-$, $F^-$ and $AcO^-$ ions but with less selectivity. It is probable that the N-H proton of imidazole unit interacts with the target anion ($X^-$) to afford an incipient N-H----X bond, and the presence of excess $X^-$ induced further N-H bond stretching eventually leading to the splitting and concomitant transfer of the N-H proton (as shown in Scheme 2). Furthermore, a very similar spectral profile of **Ru-1** in presence of hydroxyl anion ($OH^-$) with that of $CN^-$, $F^-$ and $AcO^-$ clearly suggests the ultimate deprotonation of N-H proton of coordinated pyridyl-benzimidazole ligand in **Ru-1** (Figs. S16-S17).

The deprotonation event results in a red shift of the MLCT band along with luminescence quenching. The increased electron density on the imidazole moiety strengthens both σ-donation from imidazole-N toward Ru(II) as well as π-backdonation from Ru(II) toward ligands. This promotes the charge transfer process between Ru(II) centre and the bpy (dπ⟶π*) units. On the other hand, the luminescence quenching may be attributed to the PET (photoinduced electron transfer) arising from the deprotonated imidazole moiety to the photoexcited Ru(II) in **Ru-1**. Furthermore, TRF studies suggested the sensing mechanism was due to the eventual deprotonation of the N-H group by the cyanide ion which results in two distinct species and therefore a biexponential curve could be observed.

Density functional theory (DFT) calculations have also been carried out to support the anion binding mechanism. A distorted octahedral geometry around Ru(II) centre is evident in complex **Ru-1** and its deprotonated form i.e., **Ru-1-dp**. The $Ru(bpy)_2$ unit is coordinated by the two N atoms of the pyridyl benzimidazole moiety. The related optimized structures of probe **Ru-1** and **Ru-1-dp** are presented in Fig. 6(a).

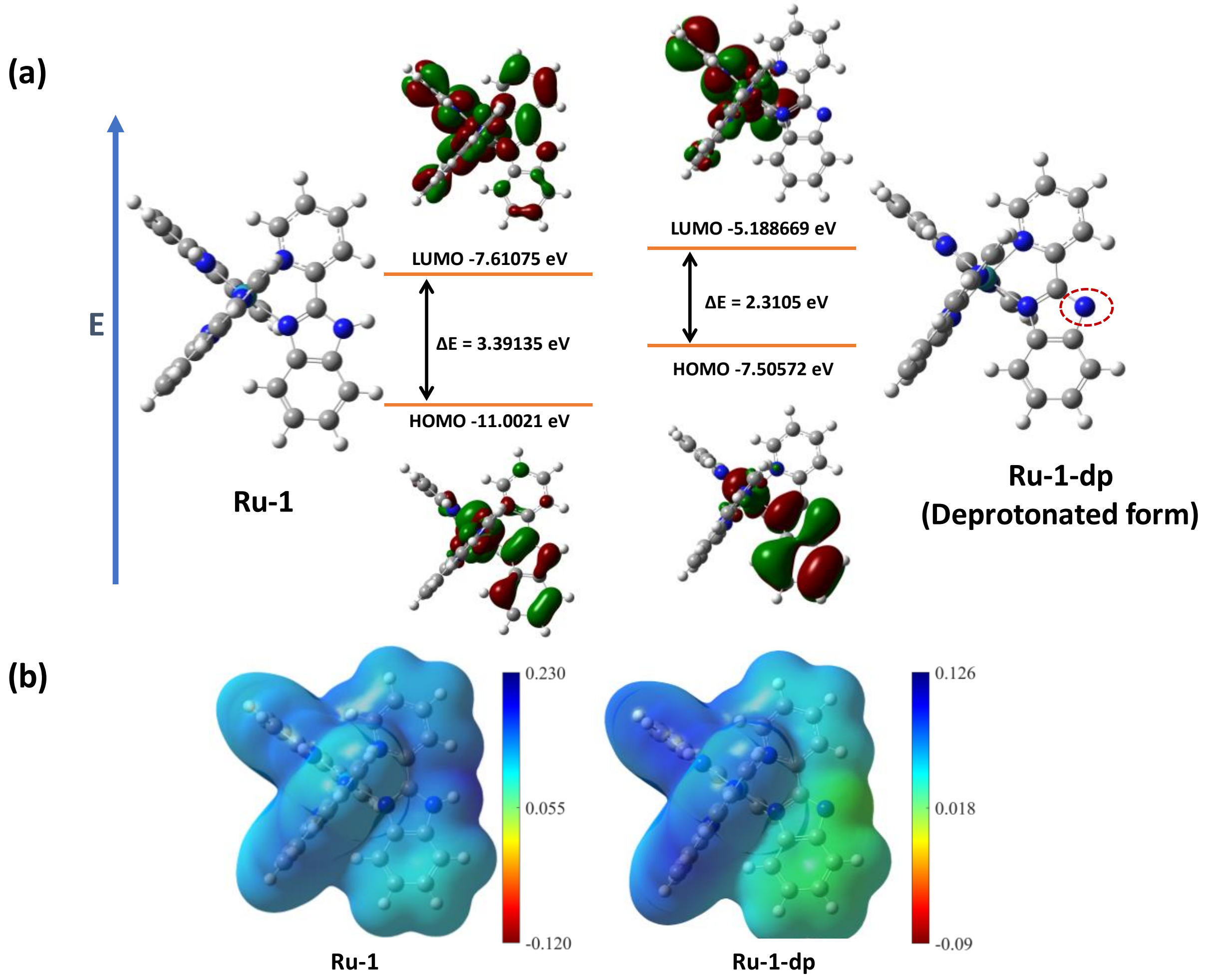

**Fig. 6**. (a) DFT optimized structures of probe **Ru-1** and its deprotonated form **Ru-1-dp** and (b) The electrostatic potential (ESP) map of the probe **Ru-1** and its deprotonated form **Ru-1-dp** on isodensity surface in ground state within the range of -0.120 (red) to + 0.230 (blue) for **Ru-1** and -0.09 (red) to +0.126 (blue) for **Ru-1-dp**.

The HOMO (i.e., highest occupied molecular orbital) of **Ru-1** was primarily composed of Ru-character with a very small contribution from the benzimidazole group of the pyridyl-benzimidazole ligand. On the other hand, its LUMO (i.e., lowest unoccupied molecular orbital) was equally localized over all three ligands. In the deprotonated form **Ru-1-dp**, the HOMO mainly consisted of the benzimidazole unit with a significant decrease in Ru contribution; however, its LUMO was mainly bpy centred. These data suggested an increased delocalization of negative charge over the benzimidazole group after deprotonation in **Ru-1-dp**. The energy gap between the frontier molecular orbitals in complex **Ru-1** was depicted to be 3.39 eV that was reduced to 2.31 eV after deprotonation process (Fig. 6(a)). The lowering in the energy gap between HOMO and LUMO of **Ru-1-dp** agreed well with the red-shifted MLCT absorption in the UV-Vis spectrum due to **Ru-1** ----$CN^-$ interaction (vide supra). The calculated ground state Ru-$N_{bpy}$ bond lengths in **Ru-1** and **Ru-1-dp** were found close to each other ranging from 2.08-2.09 Å. Notably, the Ru-$N_{imidazole}$ bond length reduced markedly from 2.10 Å to 2.07 Å after deprotonation (Tables S3 and Fig. S18). This shortening in the bond length may be ascribed to the increased synergic effect between the metal centre and nitrogen atom of the imidazole unit. The electrostatic potential (ESP) was also mapped for better visualization of the distribution of electronic charge density and its reorganization after deprotonation. The different color corresponds to the different electron density; basically, the dark blue color represents electropositive area while the bluish-green area corresponds to the electron rich site. The ESP analysis clearly revealed an increased electronic density on the benzimidazole moiety after deprotonation of **Ru-1** (Fig. 6(b)). The proposed mechanism for sensing is shown in Scheme 2.

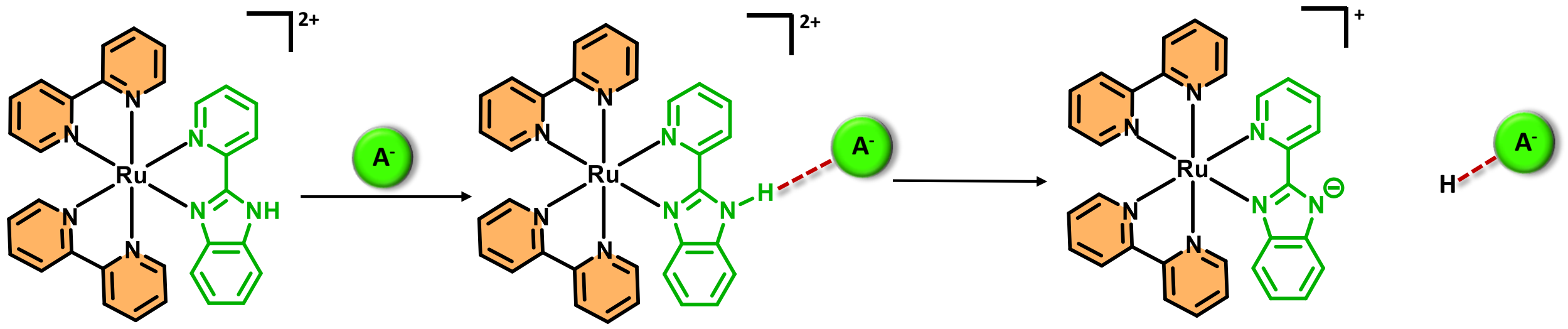

**Scheme 2**. Probable binding mode of anions ($A^-$) with probe **Ru-1** ($A^-$ = $F^-$, $AcO^-$, and $CN^-$ ions).

**Table 1**. Different mono and binuclear Ru(II)-based optical probes employed for the detection of $CN^-$ ion.

| S.No. | Probe | LoD (nM) | Real Sample | Solvent | Ref. |
|---|---|---|---|---|---|
| **(a) Reaction-based probes** | | | | | |
| 1. | 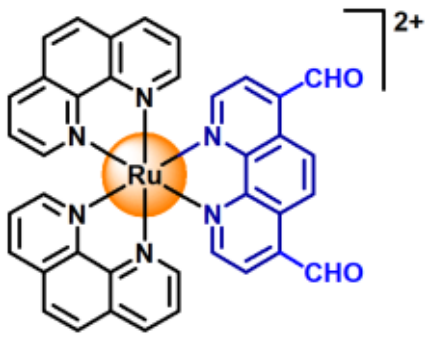 | 180 | - | $CH_3CN$ | [57] |
| 2. | 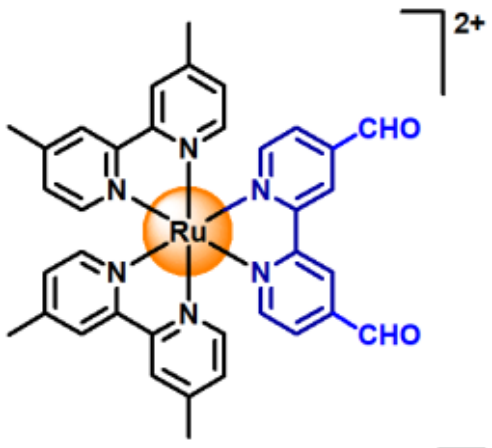 | - | - | $CH_3CN$ | [56] |
| 3. | 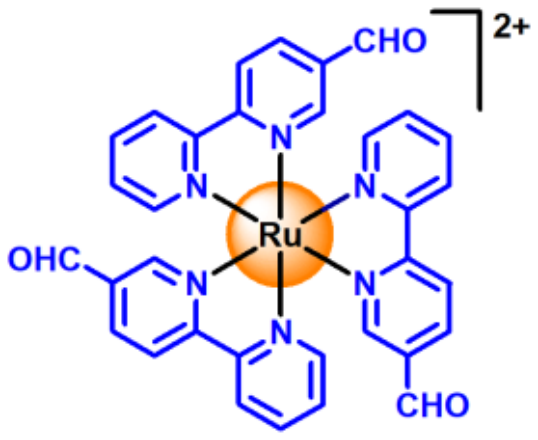 | 700 | - | $CH_3CN/H_2O$ (4:6) | [50] |
| **(b) Binding-based probes** | | | | | |
| 4. | 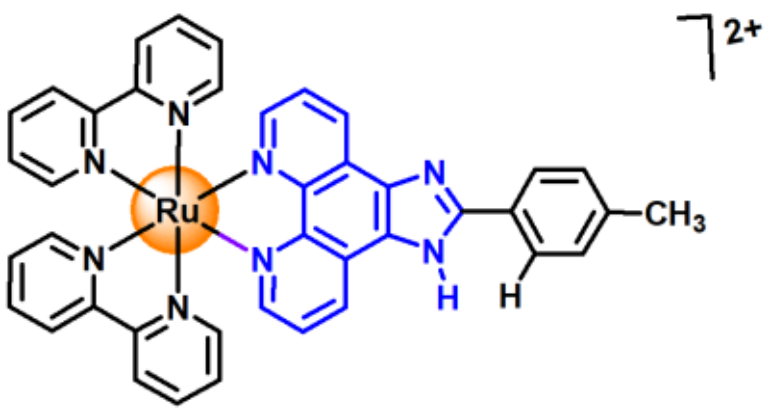 | $10^5$ | - | $H_2O$ | [53] |
| 5. | 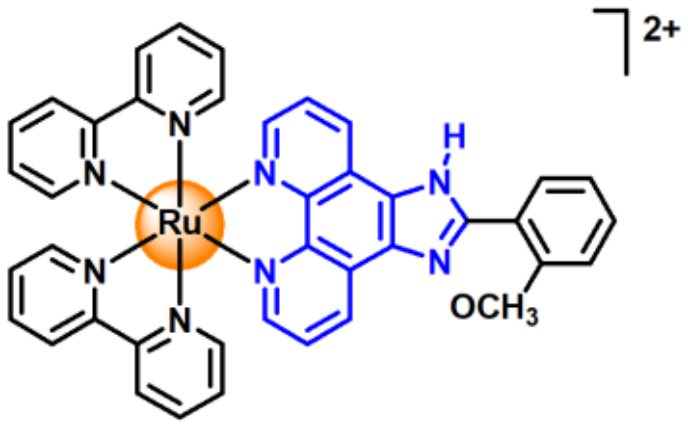 | 300 | - | $CH_3CN$ | [6] |
| 6. | 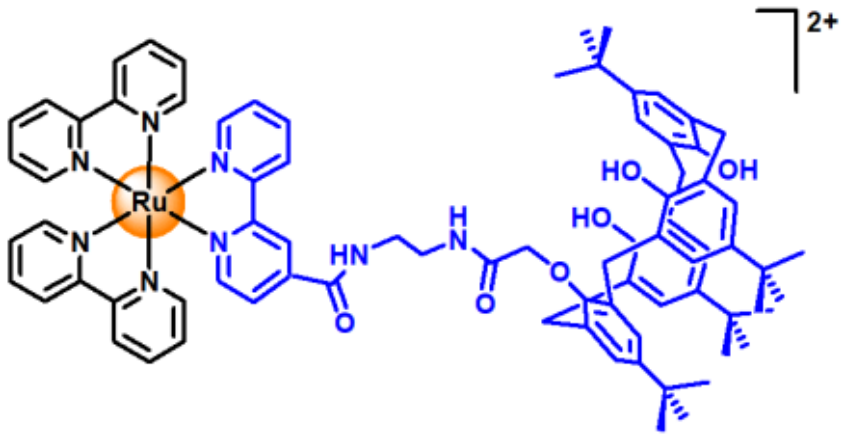 | 70 | | | [55] |
| 7. | 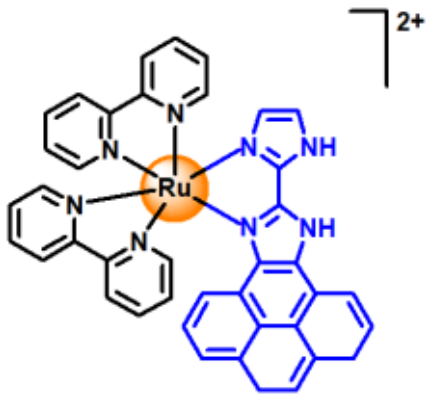 | 11.2 | Solid state | $H_2O$ | [54] |
| 8. | 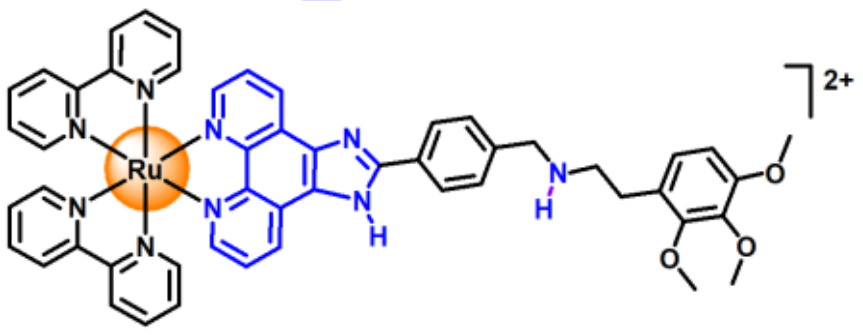 | 330 | - | $H_2O$ | [51] |
| 9 | 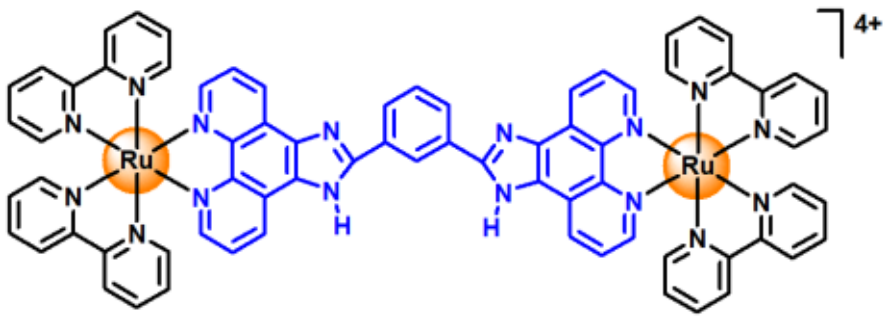 | 5000 | - | $H_2O$ | [53] |
| 10 | 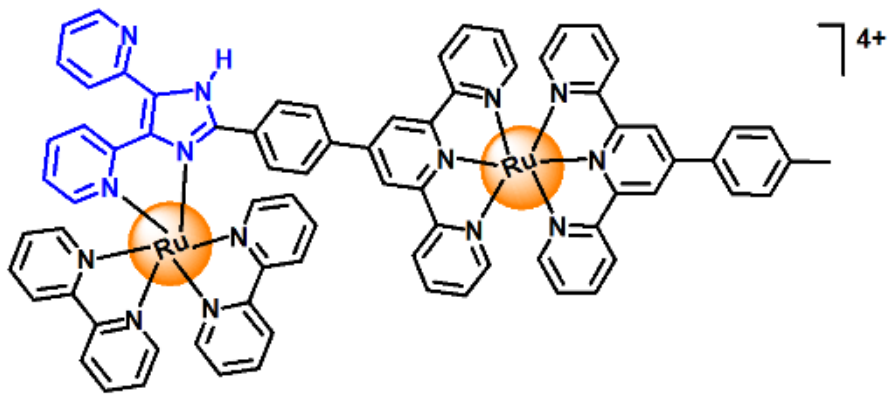 | 10 | Paper Strip | $H_2O$ | [66] |
| 11 | 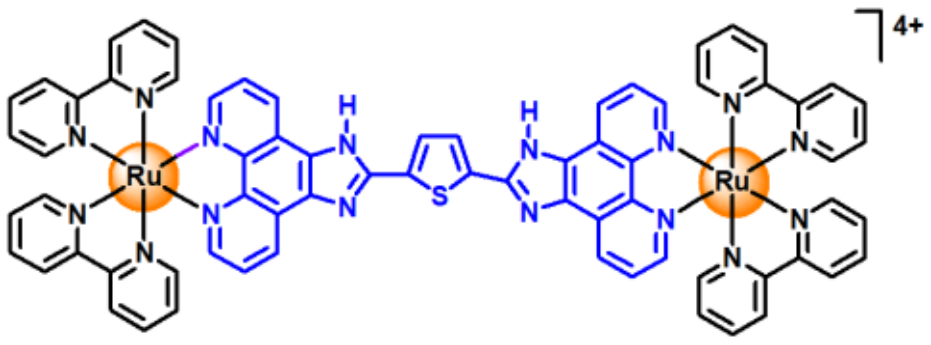 | 31 | | $H_2O$ | [52] |

| | | | | |
|---|---|---|---|---|
| **This work** | 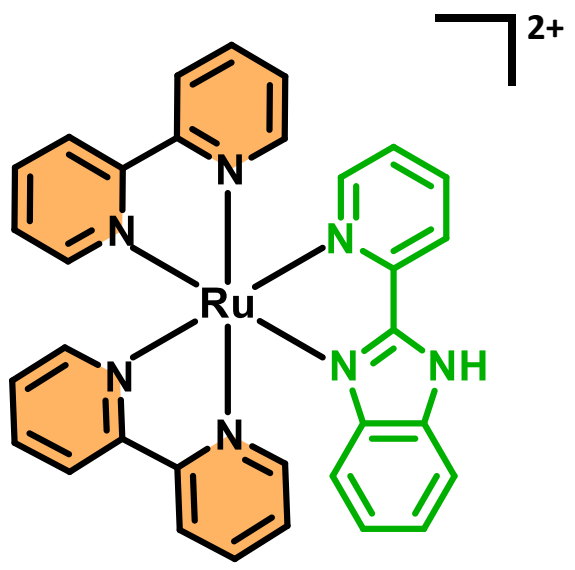 | 12.8 | Paper strip, Tap water, Distilled water, Live cell imaging | $H_2O$ |

## Conclusions

In summary, a cyanide-responsive probe with high selectivity in pure water *via* H-bonding interaction has been presented. In probe **Ru-1**, the direct coordination of imidazole moiety with metal centre leads to an enhanced acidity of N−H proton which is an important parameter to define the selectivity of **Ru-1** toward $CN^-$ ion in water. Furthermore, this probe has been exploited to assess the $CN^-$-induced emission lifetime changes in aqueous medium. The 1:1 anion binding stoichiometry has been established by Job's plot and DFT analyses. The live cell imaging assessment clearly revealed the successful imaging of $CN^-$ ions in MCF-7 cell lines. Thus, the probe **Ru-1** holds promising application prospects cyanide species analytical field, providing overall a new strategy for the future development of luminescent anion-responsive sensors.

## Acknowledgements

The authors are grateful to the Central Instrumentation Centre (CIC) at UPES for instrumentation facility. SK acknowledges Department of Science and Technology (DST), New Delhi for financial assistance as Inspire Faculty Award (IFA 15-CH213) [DST/INSPIRE/04/2015/002971]. SN sincerely acknowledges JRF scholarship provided by UPES, Dehradun.